\documentclass[draft]{agujournal}
\draftfalse

\usepackage{booktabs}
\usepackage{url}

\journalname{Space Weather}

\begin{document}

%
%


\title{Coronal Magnetic Structure of Earthbound CMEs and \textit{In situ} Comparison}

%
%




\authors{E. Palmerio\affil{1}, E. K. J. Kilpua\affil{1}, C. M\"ostl\affil{2}, V. Bothmer\affil{3}, A. W. James\affil{4}, L. M. Green\affil{4}, A. Isavnin\affil{5}, J. A. Davies\affil{6}, and R. A. Harrison\affil{6}}


\affiliation{1}{Department of Physics, University of Helsinki, P.O. Box 64, 00014 Helsinki, Finland}
\affiliation{2}{Space Research Institute, Austrian Academy of Sciences, 8042 Graz, Austria}
\affiliation{3}{Institute for Astrophysics, Georg-August-University of G\"ottingen, Friedrich-Hund-Platz 1, 37077 G\"ottingen, Germany}
\affiliation{4}{University College London, Mullard Space Science Laboratory, Holmbury St. Mary, Dorking, Surrey, RH5 6NT, UK}
\affiliation{5}{Center for Mathematical Plasma Astrophysics, Department of Mathematics, KU Leuven, 200 B, 3001 Leuven, Belgium}
\affiliation{6}{STFC-RAL Space, Rutherford Appleton Laboratory, Harwell Campus, OX11 0QX, UK}




\correspondingauthor{Erika Palmerio}{erika.palmerio@helsinki.fi}




\begin{keypoints}
\item The flux rope type is analysed and compared to signatures at the Sun and \textit{in situ} for 20 CMEs 
\item The change in the CME flux rope axis orientation from the Sun to local \textit{in situ} measurements is estimated
\item 65\% of the analysed events change their axis tilt by less than $90^{\circ}$ from the Sun to Earth
\end{keypoints}

%
%


\begin{abstract}
Predicting the magnetic field within an Earth-directed coronal mass ejection (CME) well before its arrival at Earth is one of the most important issues in space weather research. In this article, we compare the intrinsic flux rope type, \textit{i.e.} the CME orientation and handedness during eruption, with the \textit{in situ} flux rope type for 20 CME events that have been uniquely linked from Sun to Earth through heliospheric imaging. Our study shows that the intrinsic flux rope type can be estimated for CMEs originating from different source regions using a combination of indirect proxies. We find that only 20\% of the events studied match strictly between the intrinsic and \textit{in situ} flux rope types. The percentage rises to 55\% when intermediate cases (where the orientation at the Sun and/or \textit{in situ} is close to $45^{\circ}$) are considered as a match. We also determine the change in the flux rope tilt angle between the Sun and Earth. For the majority of the cases, the rotation is several tens of degrees, whilst 35\% of the events change by more than $90^{\circ}$. While occasionally the intrinsic flux rope type is a good proxy for the magnetic structure impacting Earth, our study highlights the importance of capturing the CME evolution for space weather forecasting purposes. Moreover, we emphasize that determination of the intrinsic flux rope type is a crucial input for CME forecasting models.
\end{abstract}

%
%

\section{Introduction}

Coronal mass ejections (CMEs) are large clouds of plasma and magnetic flux expelled from the Sun into the heliosphere. If directed towards Earth, they can cause significant space weather effects upon impact with the near-Earth environment. CMEs are believed to be ejected from the solar atmosphere as helical magnetic field structures known as flux ropes \citep[\textit{e.g.},][]{antiochos1999,moore2001,kliem2006,liu2008,vourlidas2014}. This flux rope structure is, however, not always observed in interplanetary space \citep[\textit{e.g.},][]{gosling1990,richardson2004b,huttunen2005}, purportedly because (1) CMEs often deform due to interactions with the ambient solar wind \citep[\textit{e.g.},][]{odstrcil1999,savani2010,manchester2017}, or with other CMEs \citep[\textit{e.g.},][]{burlaga2002,manchester2017}, (2) CMEs undergo magnetic flux erosion \citep{dasso2007,ruffenach2012}, or (3) due to the spacecraft crossing the flux rope far from its centre \citep[\textit{e.g.},][]{cane1997,jian2006,kilpua2011}. Interplanetary CMEs \citep[or ICMEs, \textit{e.g.},][]{kilpua2017b} that present, among other properties, enhanced magnetic fields, a monotonic rotation of the magnetic field direction through a large angle, small magnetic field fluctuations, and a low plasma temperature and plasma $\beta$ are often described and analysed using flux rope structures \citep[\textit{e.g.},][]{burlaga1981,rodriguez2016}.

The geoeffectivity of an ICME depends significantly on its magnetic structure, and in particular on the North--South magnetic field component (\textit{i.e.}, $B_{Z}$). A southward $B_{Z}$ will cause reconnection at the dayside magnetopause, allowing the efficient transport of solar wind energy and plasma into the magnetosphere \citep[\textit{e.g.},][]{dungey1961,gonzalez1994,pulkkinen2007}. Strong geomagnetic storms occur when the interplanetary magnetic field points strongly southward (\textit{i.e.}, $B_{Z} < -10$ nT) for more than a few hours \citep[\textit{e.g.},][]{gonzalez1987}. Due to their coherent field rotation and their tendency for enhanced magnetic fields, flux ropes are one of the key interplanetary structures that create such conditions \citep[\textit{e.g.},][]{gosling1991,huttunen2005, richardson2012,kilpua2017a}. A major goal of space weather forecasting is to be able to predict the magnitude and direction of the southward $B_{Z}$ component before the ICME arrives at Earth. The first step in achieving this aim is to understand how the magnetic field of a flux rope is organised.

\begin{figure}[ht]
\includegraphics[width=.99\linewidth]{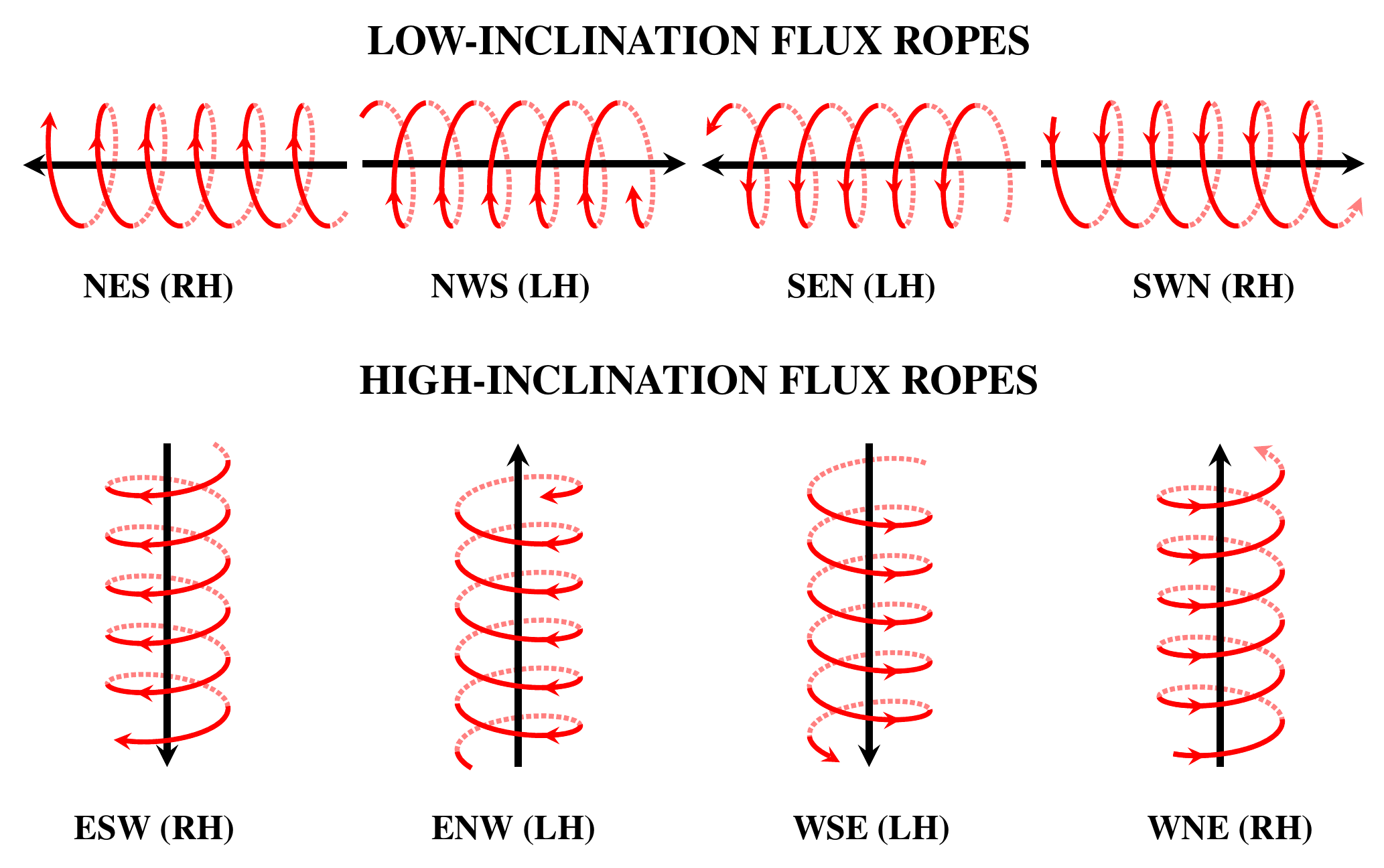}
\caption{Sketch representing the eight main flux rope types and how the helical (in red) and axial (in black) magnetic fields are related to each other for each type. Each letter describing a type represents one of the four directions (North, West, South, and East), while RH indicates right-handed and LH indicates left-handed helicity. This classification follows \citet{bothmer1998} and \citet{mulligan1998}.}
\label{fig:fr_types}
\end{figure}

The magnetic field of a flux rope can be described by two components: the helical field component, that wraps around the flux tube, and the axial field component, which runs parallel to the central axis. In addition, flux ropes can have either a left-handed or right-handed twist (chirality). Having knowledge of the flux rope chirality along with its orientation in space allows a flux rope to be classified as one of eight different ``types'', as described by \citet{bothmer1998} and \citet{mulligan1998}. Flux ropes that have their central axis more or less parallel to the ecliptic plane are called low-inclination flux ropes (in this case, the $B_{Z}$ component represents the helical field and thus its sign changes as the flux rope is crossed), while flux ropes that have their central axis more or less perpendicular to the ecliptic plane are called high-inclination flux ropes (in this case, the $B_{Z}$ component represents the axial field and thus its sign does not change). Figure \ref{fig:fr_types} shows the different flux rope types based on their chirality and orientation. There is a tendency for erupting CMEs to have negative (positive) helicity sign in the northern (southern) hemisphere. This pattern is known as the ``hemispheric helicity rule'' \citep{pevtsov2003}, but it holds only for about 60-75\% of cases \citep{pevtsov2014}.

At present, it is not possible to determine the magnetic structure of erupting flux ropes in the corona from direct observations of the magnetic field. However, several indirect proxies based on EUV, X-ray, and photospheric magnetograms have been used to estimate the ``intrinsic'' flux rope type at the time of eruption. In several studies, such proxies have been used to estimate the magnetic structure of erupting CMEs, which have been compared to \textit{in situ} observations \citep[\textit{e.g.},][]{mcallister2001,yurchyshyn2001,mostl2008,palmerio2017}. These studies have been based either on observations alone or on observations combined with theoretical and/or empirical models. In order to reconstruct the intrinsic flux rope type, the chirality sign, the axis tilt (\textit{i.e.}, its inclination to the ecliptic), and the axial direction of the magnetic field have to be known. In a force-free magnetic field configuration like a flux rope, the total magnetic helicity is conserved \citep{woltjer1958}.  Previous studies have suggested that the helicity sign, the total helicity, and the total magnetic flux of an ICME flux rope are related to those of its corresponding source region \citep[\textit{e.g.},][]{leamon2004,qiu2007,mostl2009,cho2013,hu2014,pal2017}. Hence, the property of magnetic helicity conservation can be used to assume that once the flux rope type at the Sun is determined, its chirality is maintained as the CME propagates from the Sun to Earth.        

\citet{palmerio2017} determined the magnetic structure of two CMEs both at the Sun and \textit{in situ}. The scheme presented in their work is based on the combination of multiwavelength remote-sensing observations in order to determine the chirality of the erupting flux rope and the inclination and direction of its axial field, thus reconstructing the intrinsic flux rope type. While, for the two eruptions under study, the flux rope type was the same when determined at the Sun as when measured \textit{in situ} at the Lagrange L1 point, this is not universally the case. CMEs can change their orientation due to deflections \citep[\textit{e.g.},][]{kay2013,wang2014}, rotations \citep[\textit{e.g.},][]{mostl2008,vourlidas2013,isavnin2014}, and deformations \citep[\textit{e.g.},][]{savani2010} in the corona and in interplanetary space, and this can alter the classification of the flux rope. CMEs can also change their direction, orientation, and shape due to interaction with other CMEs or corotating interaction regions \citep[CIRs,][]{lugaz2012,shen2012}. In addition, it is often difficult to predict how close a flux rope will cross Earth with respect to its nose and its central axis, and in some cases even whether a CME will encounter Earth at all \citep[\textit{e.g.},][]{mostl2014,mays2015,kay2017a}. 

In this work, we extend the study of \citet{palmerio2017}. In particular, we quantify the success of predicting flux rope types when neglecting CME evolution through a statistical analysis. The methods described by \citet{palmerio2017} provide a relatively quick and straightforward estimate of the flux rope type for space weather forecasting purposes. However, due to the potentially significant evolution of flux ropes in the corona and heliosphere through the previously described processes, the applicability of the approach has to be statistically evaluated. This is the key motivation for this study. We point out that irrespective of any direct correspondence that is found between intrinsic and \textit{in situ} flux rope types, the \citet{palmerio2017} scheme can provide a crucial input to semi-empirical CME models \citep[\textit{e.g.},][]{savani2015,savani2017,kay2016,kay2017a} or flux rope models used in numerical simulations \citep[\textit{e.g.},][]{shiota2016} that can capture the evolution. Apart from the CME evolution in the corona, changes in the axis orientation may be related to either global rotations of the whole CME body and/or to local deformations of the flux rope during its travel in the interplanetary medium, and/or to limitations of the methods used to determine the CME orientation both at the Sun and \textit{in situ}.  

This paper is organised as follows. In Section 2, we describe the spacecraft and ground-based data that we use, and also introduce the catalogue of events that we consider for this study. Then, we discuss in more detail the different methods that we have applied to determine the intrinsic flux rope type at the point of the eruption, from solar observations, and the \textit{in situ} analysis we performed. In Section 3, we apply our methods to 20 Earth-directed CMEs, by estimating the intrinsic flux rope type and comparing it to the magnetic structure measured near Earth. Finally, in Section 4, we discuss and summarize our results.

\section{Data and Methods}

\subsection{Spacecraft and Ground-based Data}

We combine various remote-sensing observations to estimate the intrinsic flux rope type of the CMEs under study and to link the interplanetary structures to their solar origins.

We use coronagraph images taken with the \textit{Large Angle Spectroscopic Coronagraph} \citep[LASCO:][]{brueckner1995} onboard the \textit{Solar and Heliospheric Observatory} \citep[SOHO:][]{domingo1995} and with the COR1 and COR2 coronagraphs that form part of the \textit{Sun Earth Connection Coronal and Heliospheric Investigation}  \citep[SECCHI:][]{howard2008} instrument package onboard the \textit{Solar Terrestrial Relations Observatory} \citep[STEREO:][]{kaiser2008}. The Heliospheric Imagers \citep[HI:][]{eyles2009} onboard STEREO are also used, primarily to connect the CMEs with their corresponding ICMEs.

We also use EUV/UV images and line-of-sight magnetograms taken with the \textit{Atmospheric Imaging Assembly} \citep[AIA:][]{lemen2012} and the \textit{Helioseismic and Magnetic Imager} \citep[HMI:][]{scherrer2012} instruments onboard the \textit{Solar Dynamics Observatory} \citep[SDO:][]{pesnell2012}. AIA takes images with a pixel size of 0.6 arcsec and a cadence of 12 seconds. HMI creates full-disc magnetograms using the 6173 \AA \, spectral line with a pixel size of 0.5 arcsec and a cadence of 45 seconds. During gaps in the AIA dataset, we use observations from the \textit{Sun-Watcher with Active Pixel System and Image Processing} \citep[SWAP:][]{berghmans2006} instrument onboard the \textit{Project for On Board Autonomy 2} (PROBA2) that images the Sun at 174 \AA \, with a cadence of one minute.

Soft X-ray data are supplied by the \textit{X-Ray Telescope} \citep[XRT:][]{golub2007} onboard \textit{Hinode} \citep[Solar-B:][]{kosugi2007}. XRT has various focal plane analysis filters, detecting X-ray emission over a wide temperature range (from 1 to 10 MK). It provides images with a pixel size of two-arcseconds.

We use H$\alpha$ (6563 \AA) observations from the \textit{Global Oscillations Network Group} (GONG) and the \textit{Global High Resolution H$\alpha$ Network} (HANET). GONG is a six-station network and HANET is a seven-station network of ground-based observatories located around the Earth to provide near-continuous observations of the Sun.

\textit{In situ} measurements are taken from the \textit{Wind} satellite. In particular, we use the data from the \textit{Wind Magnetic Fields Investigation} \citep[MFI:][]{lepping1995} and the \textit{Wind Solar Wind Experiment} \citep[SWE:][]{ogilvie1995}, which provide 60-second and 90-second resolution data, respectively.

Hourly disturbance storm time (\textit{Dst}) values are taken from the WDC for Geomagnetism, Kyoto, webpage (\url{http://wdc.kugi.kyoto-u.ac.jp/wdc/Sec3.html}). The events until 2013 are based on the final \textit{Dst} index, while those from 2014 and 2015 are based on the provisional \textit{Dst} index.

\subsection{Event Selection}

We searched the LINKed CATalogue (LINKCAT) for suitable events. LINKCAT is an output of the HELiospheric Cataloguing, Analysis and Techniques Service (HELCATS, \url{https://www.helcats-fp7.eu}) project and contains events in the timerange May 2007 to December 2013. LINKCAT connects CMEs from their solar source to their \textit{in situ} counterparts using a geometrical fitting technique based on single spacecraft data from the STEREO/HI instruments. CME tracks in HI time-elongation maps (so-called J-maps) are fitted using the Self-Similar Expansion Fitting (SSEF) method \citep{davies2012}, assuming a fixed angular half-width of $30^{\circ}$ for each CME. This yields estimates of a CME's propagation direction and radial speed. The LINKCAT catalogue consists of events where CMEs observed in HI imagery could be uniquely linked to CMEs observed in coronagraph and solar disc data and ICMEs detected \textit{in situ}.This was done by ensuring that the predicted impact of the CME based on SSEF is within $\pm 24$ hrs of the \textit{in situ arrival} time (often this is the shock arrival time). Cases where two CMEs are predicted to arrive within this window, or two ICMEs are detected within the window, are excluded, eliminating potential CME--CME interaction events. More details can be found in the online information pertaining to the catalogue (see Sources of Data and Supplementary Material). It must be kept in mind when thinking about real-time prediction that our study thus involves the down-selection to cases of a particular nature, and is based on science data. One of the ICME catalogues used to compile LINKCAT, in particular for CMEs detected towards Earth, is the Wind ICME catalogue ({\url{https://wind.nasa.gov/ICMEindex.php}). For a validation of use of the aforementioned HI-based SSEF technique to predict CME arrivals, see \citet{mostl2017}.

Since SDO is our primary spacecraft for solar observations to study the CME source region, only the LINKCAT events that arrived at Earth after May 2010 are considered. During this period, LINKCAT contains 47 Earth-impacting events. We further consider only events that present a clear flux rope \textit{in situ}, \textit{i.e.} from which we are able to estimate the flux rope type by visual inspection. We are left with 12 CME--ICME pairs.
Since LINKCAT is compiled in a semi-automated way, we also performed our own survey of on-disc CME signatures in SDO images for the events in the LINKCAT catalogue. Due to some restrictive assumptions (\textit{e.g.} $30^{\circ}$ fixed angular half width), LINKCAT does not include all possible CME--ICME pairs. 

Therefore, to find additional events for analysis we also searched other ICME catalogues, identifying ICMEs for which we could find the corresponding solar source over the period corresponding to SDO observations. In particular, we searched for additional \textit{in situ} flux ropes from the Wind ICME list and from the Near-Earth Interplanetary Coronal Mass Ejections list (\url{http://www.srl.caltech.edu/ACE/ASC/DATA/level3/icmetable2.htm}). We scanned backwards from the time at which events were observed by the HI imagers, identifying corresponding signatures in images from the COR2 and COR1 coronagraphs, and finally searched for the source on the solar disc. For those events that were not in LINKCAT, we tracked the ICME backwards in time to the Sun assuming constant speed and radial propagation, and used HI imagery to follow the CME in the heliosphere. At this stage, we utilised the HELCATS ARRival CATalogue \citep[ARRCAT,][]{mostl2017}, that lists predicted arrivals of CMEs at various spacecraft and planets using the previously described STEREO/HI SSEF fitting technique.

In the search for additional events, we also extended the time range of the data under consideration to December 2015. We identify eight additional events in this way (two due to the extension of the time range), bringing the total number of events in the study up to 20. We number the events (1--20) in chronological order of their launch times; the additional events correspond to those numbered 2, 3, 9, 10, 11, 15, 19, and 20. Event number 10 is a CME--CME interaction event in June 2012 for which the CME--ICME relation has been clarified in several previous studies \citep[\textit{e.g.},][]{kubicka2016,palmerio2017,james2017,srivastava2018}. Event number 18 is a lineup event which was also partly observed by MESSENGER, situated only a few degrees away from the Sun--Earth line \citep{moestl2018}.

\subsection{Intrinsic Flux Rope Type Determination}
\label{subsec:frtype}

As mentioned in the Introduction, in order to determine the magnetic flux rope type of an erupting CME, three parameters are needed: the chirality, the axis orientation, and the axial field direction. The chirality can be inferred from several multi-wavelength proxies: magnetic tongues \citep{lopezfuentes2000,luoni2011}, X-ray and/or EUV sigmoids and/or sheared arcades \citep[\textit{e.g.},][]{rust1996,canfield1999,green2007}, the skew of coronal arcades \citep{mcallister1998,martin2012}, flare ribbons \citep{demoulin1996}, and filament details \citep{martin1994,martin1996,chae2000}. For a detailed description of these helicity proxies, see \citet{palmerio2017}.

The inclination of the flux rope axis with respect to the ecliptic, $\tau$, is taken to be the average of the orientation of the polarity inversion line \citep[PIL,][]{marubashi2015} and the orientation of the post-eruption arcades \citep[PEAs,][]{yurchyshyn2008}, in the range $[-90^{\circ},90^{\circ}]$. The tilt angle $\tau$ is measured from the solar East, and assumes a positive (negative) value if the acute angle to the ecliptic is to the North (South). For source regions where the PIL can easily be approximated as a straight line (\textit{e.g.}, quiet Sun and magnetically simple active regions), we determine the PIL orientation by eye, \textit{i.e.} we determine the location where the polarity of the magnetic field reverses, and approximate it as a straight line. When the PIL is more curved and/or complex, we smooth the data over square bins containing variable numbers of pixels, overplot the locations where $B_{r} = 0$, and then estimate the orientation of the resulting PIL. For source regions located between $\pm 30^{\circ}$ in longitude on the solar disc, we use HMI line-of-sight data. For source regions located closer to the limb, in order to reduce the projection effects, we use \textit{Space-weather HMI Active Region Patch} \citep[SHARP:][]{bobra2014} data, derived with the series \textit{hmi.sharp\_cea\_720s} where the vector \textbf{B} has been remapped onto a Lambert Cylindrical Equal-Area (CEA) projection. Similarly, the orientation of the PEAs is determined by eye for source regions located between $\pm 30^{\circ}$ in longitude on the solar disc, while for regions located nearer the limb, we correct the projection effects by first converting two points on the arcade axis from Helioprojective-Cartesian to Heliographic coordinates. Then, we apply to the axis the vector rotation operator ``rotate'', defined as
\begin{equation}
\text{rotate}(\mathbf{\hat{v}},\mathbf{\hat{a}},\gamma)=\mathbf{\hat{v}}\cos{\gamma}+(\mathbf{\hat{v}}\cdot\mathbf{\hat{a}})(1-\cos{\gamma})\mathbf{\hat{a}}+[\mathbf{\hat{a}}\times\mathbf{\hat{v}}]\sin{\gamma} \, ,
\end{equation}
which rotates the arcade axis, $\mathbf{\hat{v}}$, counterclockwise around its median, $\mathbf{\hat{a}}$, by a tilt angle, $\gamma$  \citep{isavnin2013}. We rotate the axis until it becomes parallel to the ecliptic. The total rotation corresponds to the unprojected tilt of the arcade's axis. 

For some events, we could only estimate the orientation of the axis from the PIL direction, because PEAs were either too short or not visible. When we have obtained the average orientation between PIL and PEAs, we assume:
\begin{enumerate}
\item $0^{\circ}\leq |\tau| <35^{\circ} \Rightarrow$ low-inclination flux rope
\item $35^{\circ}\leq |\tau |\leq 55^{\circ} \Rightarrow$ intermediate flux rope
\item $55^{\circ}< |\tau| \leq 90^{\circ} \Rightarrow$ high-inclination flux rope
\end{enumerate}

Finally, we check the direction of the axial field by looking at coronal dimmings in EUV difference images and identifying in which magnetic polarities they are rooted. Then, the magnetic field direction is defined from the positive polarity to the negative one. When the three parameters are known, we can reconstruct the flux rope type at the point of the eruption. 

\subsection{\textit{In situ} Flux Rope Type Identification}
\label{subsec:insitu}

The CME flux rope type at the time of the eruption is compared to the magnetic configuration of the corresponding ICME. First, we analyse, by eye, the magnetic field components of the ICME observed \textit{in situ} in both Cartesian ($B_{x}$, $B_{y}$, $B_{z}$) and angular ($B_{\theta}$, $B_{\phi}$) geocentric solar ecliptic (GSE) coordinates, and make a first estimate of the type of the \textit{in situ} flux rope.

We then apply minimum variance analysis \citep[MVA,][]{sonnerup1967} to the \textit{in situ} measurements during the flux rope interval, to estimate the orientation of the flux rope axis (latitude, $\theta_{\text{MVA}}$, and longitude, $\phi_{\text{MVA}}$) and obtain its helicity sign. The latter is done by inspection of the direction of the magnetic field rotation in the intermediate-to-maximum plane. The flux rope axis corresponds to the MVA intermediate variance direction, where $\theta_{\text{MVA}} = 90^{\circ}$ is defined as being northward and $\phi_{\text{MVA}} = 90^{\circ}$ is defined as being eastward. We apply the MVA to 20-minute averaged magnetic field data. We also consider the intermediate-to-mininum eigenvalue ratio ($\lambda_{2}/\lambda_{3}$) resulting from MVA. MVA can be considered most reliable when $\lambda_{2}/\lambda_{3} \geq 2$ \citep[\textit{e.g.},][]{lepping1980,bothmer1998,huttunen2005}.

As a proxy for the spacecraft crossing distance from the flux rope central axis (or impact parameter), we calculate the ratio of the minimum variance direction to the total magnetic field in the MVA frame \citep{gulisano2007,demoulin2009}, $\langle|B_{\min}|\rangle/\langle B\rangle$. We average the quantities along the whole flux rope interval. A higher ratio indicates that the flux rope has been crossed progressively farther from its central axis, and it implies that the bias in the flux rope orientation is larger.

As a proxy for the spacecraft crossing distance from the nose of the flux rope, we calculate the location angle, L, defined by \citet{janvier2013} as
\begin{equation}
\text{sin}\,\text{L} = \text{cos}\,\theta_{\text{MVA}}\,\text{cos}\,\phi_{\text{MVA}} \, .
\end{equation}
The location angle ranges from $\text{L} \approx -90^{\circ}$ in one leg, through $\text{L} \approx 0^{\circ}$ at the nose, to $\text{L}\approx 90^{\circ}$ in the other leg.

Finally, we check the minimum value of the (\textit{Dst}) index related to each event. We only quote the events for which \textit{Dst}$_{\text{min}} < -50$. We consider those events with $-50$ > \textit{Dst}$_{\text{min}} > -100$ as moderate storms, and those events for which \textit{Dst}$_{\text{min}} \leq -100$ as major storms.

\subsection{Orientation Angles}

The next step is to compare the orientations of the CME axis at the Sun and \textit{in situ}. Regarding the former, we convert the tilt angle, $\tau$, into the orientation angle, $\alpha_{\text{SUN}}$, that lies within the range $[-180^{\circ},180^{\circ}]$. $\alpha_{\text{SUN}}$ is derived from $\tau$ by taking into account in which direction the flux rope axial field is pointing, that was previously estimated from coronal dimmings (see Section \ref{subsec:frtype}). The orientation angle is calculated from the positive East direction, clockwise for positive values and counterclockwise for negative values. \citet{yurchyshyn2008} determined the flux rope orientation of 25 CME events at the Sun from PEAs only, and estimated that the PEAs angles were measured with accuracy  $\pm 10^{\circ}$ for 19 events, and $\pm 90^{\circ}$ for the remaining six. Since our flux rope orientations at the Sun are determined by a combination of PIL and PEAs,  we estimate that the tilt angles were measured with an accuracy between $\pm 5^{\circ}$ (for the cases where PIL and PEAs had an almost identical orientation) and $\pm 15^{\circ}$--$20^{\circ}$ (for the cases when we could only use the PIL direction, or the PIL and PEAs directions had a larger angular separation).

Regarding the orientation of the \textit{in situ} flux rope at the Lagrange L1 point, we project the axis resulting from the MVA analysis onto a 2D plane that corresponds to the YZ-plane in GSE coordinates. We then measure the \textit{in situ} clock angle orientation, $\alpha_{\text{L1}}$, within the range $[-180^{\circ},180^{\circ}]$ as for $\alpha_{\text{SUN}}$. The MVA fittings introduce an error of $\pm 5^{\circ}$--$10^{\circ}$ when the spacecraft crosses the flux rope axis approximately perpendicularly. However, for crossings that are progressively farther from the central axis, the error on the estimated flux rope axis orientation can be up to $\pm 90^{\circ}$ \citep{owens2012}. In particular, \citet{gulisano2007} studied in detail the bias introduced in MVA fittings for flux ropes. They found that $\theta_{\text{MVA}}$ is best determined for flux ropes that have their axis close to the ecliptic plane and nearly perpendicular to the Sun-Earth line. Moreover, the angle $\eta$ between the true flux rope orientation and the MVA-generated one is $\eta\approx 3^{\circ}$ for a spacecraft crossing a cloud within 30\% of its radius, and $\eta \lesssim 20^{\circ}$ for an impact parameter as high as 90\% of the flux rope radius. One of the main issues in flux rope fittings with MVA is, therefore, the fact that the impact parameter is unknown.

\section{Results} 
\label{sec:results}

The source regions of the 20 analysed CMEs have the following properties:
\begin{itemize}
\item 10 (50\%) CMEs erupted from the Northern hemisphere and 10 (50\%) from the Southern hemisphere.
\item 14 (70\%) CMEs erupted from an active region, two (10\%) from between two active regions, and four (20\%) from a quiet Sun filament. 
\item 18 (90\%) of source regions followed the hemispheric helicity rule, while two (10\%) did not. 
\end{itemize}

Table \ref{tab:helicity_proxies} shows which helicity sign proxies were used for each event. The proxy that we could use the most (applicable to 18 events or 90\%) is the skew of the coronal arcades. This is not surprising, considering that most CMEs are associated with arcades before and/or after an eruption. These arcades can either be the coronal loops that overlie the eruptive structure or arcades that form under the CME due to magnetic reconnection after it is ejected. In a few cases, however, the arcade skew was not clear enough to be used as a helicity proxy. Clear S-shaped features were found for 14 (70\%) events. We consider here both sheared arcades and sigmoids, which are structures that can be seen in X-ray and sometimes also in EUV. Sheared arcades are multi-loop systems, while sigmoids are single-loop S-shaped structures \citep[\textit{e.g.},][]{green2007}. Sigmoids and arcades that have forward (reverse) S-shape indicate positive (negative) helicity.  Another popular chirality proxy is the use of flare ribbons. We were able to use this proxy for 11 (55\%) events. It is worth remarking that flare ribbons can be used to estimate the helicity sign of a CME and its source region if they form clear J-shapes, where a forward (reverse) J indicates positive (negative) helicity, or if they are significantly shifted along the PIL. A filament association was found for 12 (60\%) CMEs, and  for all of these we were able to use filament characteristics to estimate the chirality. We analysed both H$\alpha$ details, \textit{i.e.} filament spine shape and barbs, and EUV details, \textit{i.e.} the crossings of dark and bright threads. H$\alpha$ characteristics are mostly visible in quiet Sun filaments, while absorption and emission threads are mostly visible in active region filaments. Only for one event (Event 9) were we able to analyse the filament successfully both in H$\alpha$ and EUV. The least applicable proxy involves the use of magnetic tongues. We were only able to apply this technique to three (15\%) events. This is expected, as magnetic tongues are only visible in emerging active regions. Finally, we emphasize that, for each analysed event, all helicity sign proxies agree with one another.

\begin{table}
\caption{A summary of the chirality and shear determinations used for each of the CMEs studied. The table shows, from left to right: event number, Solar Object Locator (SOL), eruption time rounded to the nearest hour, and the chirality made possible due to the presence of magnetic tongues, proxies visible in H$\alpha$ related to the chirality of a filament, absorption and emission filament threads visible in EUV, S-shaped structure (sheared arcade or sigmoid) in EUV or X-rays, skew of coronal loops, and J-shaped flare ribbons.}
\label{tab:helicity_proxies}
\begin{tabular}{l @{\hskip 0.1in}l c c c  c c c c}
\toprule
\# & Eruption  & Tongues & H$\alpha$-fil & EUV-fil & S-shape & Skew & Ribbons\\
\midrule
1 & SOL2010-05-23, 17 UT  & - & LH & - & - & LH & - \\
2 & SOL2011-03-25, 06 UT & - & - & - & - & RH & - \\
3 & SOL2011-06-02, 07 UT & - & - & RH & RH & RH & - \\
4 & SOL2011-09-13, 22 UT & - & - & - & LH & LH & - \\
5 & SOL2011-10-22, 01 UT & - & LH & -  & - & LH & LH \\
6 & SOL2012-01-19, 14 UT & - & - & -  & - & LH & LH \\
7 & SOL2012-03-10, 17 UT & LH & - & LH & LH & LH & LH \\
8 & SOL2012-03-13, 17 UT & LH & - & LH & LH & LH & LH \\
9 & SOL2012-05-11, 23 UT & - & RH & RH & RH & RH & RH \\
10 & SOL2012-06-14, 13 UT & RH & - & - & RH & RH & - \\
11 & SOL2012-07-04, 17 UT & - & - & LH & LH & LH & LH \\
12 & SOL2012-07-12, 16 UT & - & - & RH & RH & RH & RH \\
13 & SOL2012-10-05, 00 UT & - & - & - & RH & - & - \\
14 & SOL2012-10-08, 21 UT & - & - & - & LH & LH & - \\
15 & SOL2012-10-27, 12 UT & - & - & - & - & RH & - \\
16 & SOL2013-01-13, 00 UT & - & - & - & RH & RH & - \\
17 & SOL2013-04-11, 07 UT & - & - & LH & LH & - & LH \\
18 & SOL2013-07-09, 14 UT & - & LH & - & LH & LH & LH \\
19 & SOL2014-08-15, 16 UT & - & RH & - & - & RH & RH \\
20 & SOL2015-12-16, 08 UT & - & - & RH & RH & RH & RH \\
\bottomrule
\end{tabular}
\end{table}          

Table \ref{tab:CME_types} lists the estimated flux rope types at the Sun and Table \ref{tab:ICME_types} the local flux rope types observed \textit{in situ}. We note that the chirality of the intrinsic flux rope and \textit{in situ} flux rope matched for all 20 events, including the two events that did not follow the hemispheric helicity rule. This result is expected, as the helicity sign should be preserved during interplanetary propagation, and it also gives further confirmation that our indirect helicity proxies derived from solar observations are correct. For two events (numbers 6 and 16), the MVA intermediate-to-medium eigenvalue ratio was $\lambda_{2}/\lambda_{3} < 2$, but the flux rope orientation resulting from MVA agreed with the flux rope type obtained from visual inspection.

\begin{sidewaystable}
\caption{The results of the analysis of the magnetic structure of the flux rope on the Sun. The table shows, from left to right: event number, Solar Object Locator (SOL), eruption time rounded to the nearest hour, CME source (QS: Quiet Sun, NH: Northern Hemisphere, SH: Southern Hemisphere, AR: Active Region), chirality of the erupting flux rope, whether the chirality follows the hemispheric helicity rule (HHR), inclination of the PIL, inclination of the PEAs, average tilt of the axis with respect to the ecliptic plane, direction of the axial field, and erupting flux rope type.} 
\label{tab:CME_types}
\begin{tabular}{l @{\hskip 0.2in}l c c c c c c c c c}
\toprule
\# & CME & & & & & & & & & \\
\cmidrule(r){2-11} 
& SOL & Eruption time  & Source & Chirality & HHR & PIL & PEAs & Tilt & Axial field & FR type \\
\midrule
1 & SOL2010-05-23 & 17 UT & QS, NH & LH & Yes & $38^{\circ}$ & $50^{\circ}$ & $44^{\circ}$& Southwest & WSE/NWS \\
2 & SOL2011-03-25 & 06 UT & AR 11176 & RH & Yes & $-86^{\circ}$ & -- & $-86^{\circ}$& South & ESW \\
3 & SOL2011-06-02 & 07 UT & AR 11226/11227 & RH & Yes & $-45^{\circ}$ & -- & $-45^{\circ}$& Northwest & WNE/SWN \\
4 & SOL2011-09-13 & 22 UT & AR 11289 & LH & Yes & $40^{\circ}$ & $40^{\circ}$ & $40^{\circ}$& Southwest & WSE/NWS \\
5 & SOL2011-10-22 & 01 UT & QS, NH & LH & Yes & $32^{\circ}$ & $34^{\circ}$ & $33^{\circ}$& East & SEN \\
6 & SOL2012-01-19 & 14 UT & AR 11402 & LH & Yes & $-80^{\circ}$ & $-88^{\circ}$ & $-84^{\circ}$& South & WSE \\
7 & SOL2012-03-10 & 17 UT & AR 11429 & LH & Yes & $26^{\circ}$ & $38^{\circ}$ & $32^{\circ}$ & East & SEN \\
8 & SOL2012-03-13 & 17 UT & AR 11429 & LH & Yes & $40^{\circ}$ & $46^{\circ}$ & $43^{\circ}$& Northeast & ENW/SEN \\
9 & SOL2012-05-11 & 23 UT & small AR, SH & RH & Yes & $-65^{\circ}$ & $-65^{\circ}$ & $-65^{\circ}$ & South & ESW \\
10 & SOL2012-06-14 & 13 UT & AR 11504 & RH & Yes & $-30^{\circ}$ & -- & $-30^{\circ}$& East & NES \\
11 & SOL2012-07-04 & 17 UT & AR 11513 & LH & Yes & $46^{\circ}$ & $36^{\circ}$ & $41^{\circ}$& Southwest & WSE/NWS \\
12 & SOL2012-07-12 & 16 UT & AR 11520 & RH & Yes & $-30^{\circ}$ & $-14^{\circ}$ & $-22^{\circ}$& East & NES \\
13 & SOL2012-10-05 & 00 UT & AR 11582/11584 & RH & Yes & $-73^{\circ}$ & -- & $-73^{\circ}$& South & ESW \\
14 & SOL2012-10-08 & 21 UT & AR 11585 & LH & No & $47^{\circ}$ & -- & $47^{\circ}$& Northeast & ENW/SEN \\
15 & SOL2012-10-27 & 12 UT & AR 11598 & RH & Yes & $-50^{\circ}$ & -- & $-50^{\circ}$& Southeast & ESW/NES \\
16 & SOL2013-01-13 & 00 UT & AR 11654 & RH & No & $-88^{\circ}$ & -- & $-88^{\circ}$& North & WNE \\
17 & SOL2013-04-11 & 07 UT & AR 11719 & LH & Yes & $60^{\circ}$ & $50^{\circ}$ & $55^{\circ}$& Southwest & WSE/NWS \\
18 & SOL2013-07-09 & 14 UT & QS, NH & LH & Yes & $47^{\circ}$ & $53^{\circ}$ & $50^{\circ}$& Southwest & WSE/NWS \\
19 & SOL2014-08-15 & 16 UT & QS, SH & RH & Yes & $82^{\circ}$ & $70^{\circ}$ & $76^{\circ}$& North & WNE \\
20 & SOL2015-12-16 & 08 UT & AR 12468 & RH & Yes & $-32^{\circ}$ & $-24^{\circ}$ & $-28^{\circ}$ & East & NES \\
\bottomrule
\end{tabular}
\end{sidewaystable}  

\begin{sidewaystable}
\caption{The results of the analysis of the magnetic structure of the flux rope \textit{in situ}. The table shows, from left to right: arrival time of the ICME flux rope leading edge, time of the ICME flux rope trailing edge, chirality of the \textit{in situ} flux rope, flux rope axis from MVA in the form (latitude, longitude), MVA intermediate-to-minimum eigenvalue ratio, ratio of the MVA minimum variance component to the total magnetic field (proxy for the impact parameter or crossing distance from the ICME axis), location angle (proxy for the crossing distance from the ICME nose), minimum \textit{Dst} index value (only for events \textit{Dst} < -50), and \textit{in situ} flux rope type from visual inspection.}
\label{tab:ICME_types}
\begin{tabular}{l @{\hskip 0.2in}l c c c c c c c c}
\toprule
\# & ICME & & & & & &\\
\cmidrule(r){2-10}
& Leading Edge & Trailing Edge & Chirality & MVA Axis & $\lambda_{2}/\lambda_{3} $ & $\langle|B_{\min}|\rangle/\langle B\rangle$ & L-angle &\textit{Dst}$_{\text{min}}$ & FR type\\
\midrule
1 & 2010-05-28, 19:10 & 2010-05-29, 16:50 & LH & ($-59^{\circ}$, $234^{\circ}$) & 17.9 & 0.08 & $-18^{\circ}$ & $-80$ & WSE\\
2 & 2011-03-30, 00:25 & 2011-04-01, 15:05 & RH & ($17^{\circ}$, $119^{\circ}$) & 2.9 & 0.13 & $-28^{\circ}$ & -- & NES\\
3 & 2011-06-05, 01:58 & 2011-06-05, 08:55 & RH & ($68^{\circ}$, $135^{\circ}$) & 3.9 & 0.10 & $-15^{\circ}$ & -- & WNE\\
4 & 2011-09-17, 15:38 & 2011-09-18, 08:46 & LH & ($46^{\circ}$, $70^{\circ}$) & 4.5 & 0.19 & $14^{\circ}$ & $-72$ & ENW/SEN\\
5 & 2011-10-25, 00:30 & 2011-10-25, 17:09 & LH & ($74^{\circ}$, $56^{\circ}$) & 2.7 & 0.22 & $9^{\circ}$ & $-147$ & ENW\\
6 & 2012-01-22, 11:40 & 2012-01-23, 07:55 & LH & ($-49^{\circ}$, $263^{\circ}$) & 1.9 & 0.48 & $-5^{\circ}$ & $-71$ & NWS/WSE\\
7 & 2012-03-12, 10:05 & 2012-03-12, 14:55 & LH & ($-16^{\circ}$, $35^{\circ}$) & 2.6 & 0.45 & $52^{\circ}$ & $-64$ & SEN\\
8 & 2012-03-15, 15:52 & 2012-03-16, 14:06 & LH & ($65^{\circ}$, $105^{\circ}$) & 2.2 & 0.39 & $-6^{\circ}$ & $-88$ & ENW\\
9 & 2012-05-16, 16:00 & 2012-05-17, 22:20 & RH & ($46^{\circ}$, $271^{\circ}$) & 27.9 & 0.17 & $1^{\circ}$ & -- & SWN/WNE\\
10 & 2012-06-16, 22:10 & 2012-06-17, 12:30 & RH & ($-28^{\circ}$, $99^{\circ}$) & 19.3 & 0.10 & $-8^{\circ}$ & $-86$ & NES\\
11 & 2012-07-08, 23:48 & 2012-07-09, 20:56 & LH & ($-50^{\circ}$, $340^{\circ}$) & 5.2 & 0.38 & $37^{\circ}$ & $-78$ & WSE\\
12 & 2012-07-15, 06:16 & 2012-07-16, 14:33 & RH & ($-4^{\circ}$, $305^{\circ}$) & 5.8 & 0.57 & $35^{\circ}$ & $-139$ & ESW\\
13 & 2012-10-08, 17:15 & 2012-10-09, 13:34 & RH & ($-66^{\circ}$, $258^{\circ}$) & 8.9 & 0.30 & $-5^{\circ}$ & $-109$ & ESW\\
14 & 2012-10-12, 15:50 & 2012-10-13, 09:42 & LH & ($-60^{\circ}$, $247^{\circ}$) & 10.6 & 0.38 & $-11^{\circ}$ & $-90$ & WSE\\
15 & 2012-10-31, 23:32 & 2012-11-02, 02:30 & RH & ($-68^{\circ}$, $49^{\circ}$) & 51.2 & 0.12 & $14^{\circ}$ & $-65$ & ESW\\
16 & 2013-01-17, 16:13 & 2013-01-18, 11:48 & RH & ($18^{\circ}$, $250^{\circ}$) & 1.4 & 0.16 & $-19^{\circ}$ & $-52$ & SWN\\
17 & 2013-04-14, 16:10 & 2013-04-15, 20:42 & LH & ($62^{\circ}$, $337^{\circ}$) & 6.4 & 0.17 & $26^{\circ}$ & -- & ENW\\
18 & 2013-07-13, 04:55 & 2013-07-14, 23:30 & LH & ($-10^{\circ}$, $286^{\circ}$) & 13.5 & 0.08 & $16^{\circ}$ & $-81$ & NWS\\
19 & 2014-08-19, 17:25 & 2014-08-21, 00:07 & RH & ($65^{\circ}$, $314^{\circ}$) & 48.5 & 0.07 & $17^{\circ}$ & -- & WNE\\
20 & 2015-12-20, 02:55 & 2015-12-21, 20:25 & RH & ($-30^{\circ}$, $221^{\circ}$) & 3.8 & 0.43 & $-41^{\circ}$ & $-155$ & ESW\\
\bottomrule
\end{tabular}
\end{sidewaystable} 

The flux rope types (Figure \ref{fig:fr_types}) at the Sun and \textit{in situ} match strictly for only four (20\%) of the 20 events (Events 7, 10, 13, and 19). Figure \ref{fig:event10} gives an example of such an event (Event 10). Figure \ref{fig:event10}a shows an SDO/HMI line-of-sight (LOS) magnetogram approximately two days before the eruption, when the active region was emerging, revealing the presence of right-handed magnetic tongues. Figure \ref{fig:event10}b shows a sigmoid seen in EUV that also suggests positive helicity. Another helicity proxy that we used for this event is the skew of arcade loops (not shown). The orientation of the neutral line is shown in panel \ref{fig:event10}c and has a tilt $\tau = -30^{\circ}$. The axial field points to the East. As explained in Section \ref{subsec:frtype}, this can be deduced from the locations of the EUV dimmings associated with the flux rope footpoints that are overlaid with SDO/HMI magnetogram data (Figure \ref{fig:event10}d). The previously described solar observations yield a NES-type flux rope. \textit{In situ} observations are shown on the right-hand side of Figure \ref{fig:event10}. The ICME was preceded by a shock (red line), and the flux rope (bounded between the pair of blue lines) is clearly identified from the enhanced magnetic field and smooth rotation of the field direction. MVA yields the axis of tilt $-28^{\circ}$, the fact that the field at the axis points to the East, and that the chirality is right-handed. Hence, the flux rope type \textit{in situ} is also NES, and the axis tilts at the Sun and \textit{in situ} are almost identical. 

\begin{figure}[ht]
\includegraphics[width=.99\textwidth]{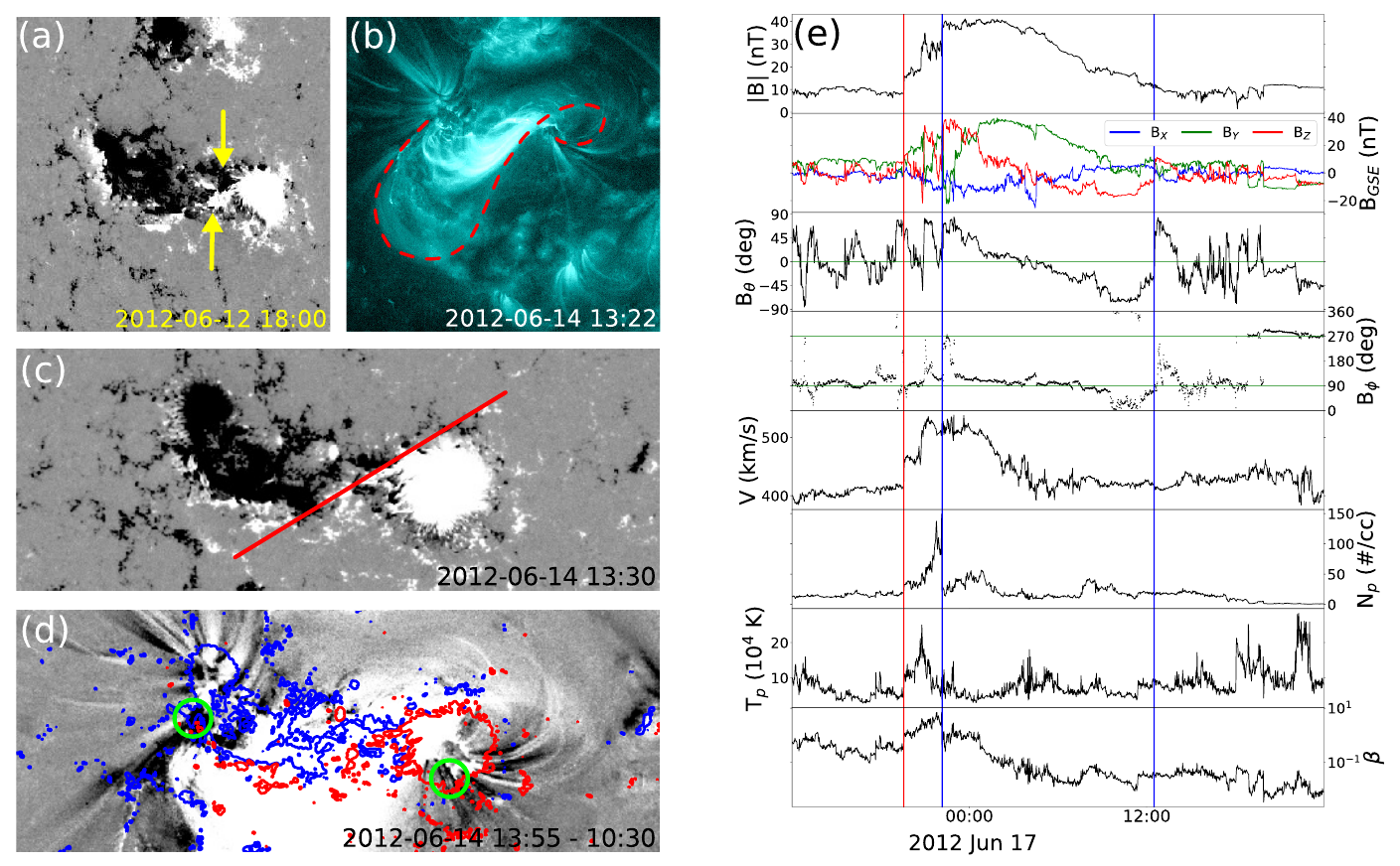}
\caption{Event 10, which is found to be a NES-type both at the Sun and \textit{in situ}. (a) Magnetic tongues as seen in an SDO/HMI magnetogram (saturated at $\pm 200$ G) that show positive chirality. (b) Forward-S sigmoid as seen by SDO/AIA 131 \AA \, that indicates a right-handed flux rope. (c) HMI magnetogram (saturated at $\pm 200$ G) showing the PIL approximated as a straight line (in red). (d) Base-difference AIA image in 131 \AA \, saturated at $\pm 70$ DN s$^{-1}$ pixel$^{-1}$ and overlaid with HMI magnetogram contours saturated at $\pm 200$ G (blue = negative polarity, red = positive polarity). The dimming regions (signatures of the flux rope footpoints) have been circled in green. (e) The ICME as observed in situ by \textit{Wind}. The red line indicates the arrival of the IP shock, while the blue lines indicate the leading and trailing edges of the flux rope. The parameters shown are, from top to bottom: magnetic field magnitude, magnetic field components in GSE cartesian coordinates, $\theta$ and $\phi$ components of the magnetic field in GSE angular coordinates, solar wind speed, proton density, proton temperature, and plasma $\beta$.}
\label{fig:event10}
\end{figure}

We emphasize that, for a significant fraction of events (nine or 45\%), the tilt angle at the Sun and/or the latitude of the \textit{in situ} flux rope axis was close to $45^{\circ}$. For such cases, considering the possible errors, one cannot distinguish between low and high-inclination flux rope types. We categorise these cases as intermediate-inclination events (see Section \ref{subsec:frtype}). An example of such an event is Event 18 (Figure \ref{fig:event18}). The left-handed chirality of this event could be determined at the Sun from H$\alpha$ filament details, arcade skew, flare ribbons, and S-shape of the filament seen in EUV. The average between the PIL tilt (Figure \ref{fig:event18}c) and the PEAs' tilt (not shown) gives a tilt angle at the Sun of $50^{\circ}$. The axial field points to the Southwest, \textit{i.e.} the possible intrinsic flux rope types are either a high-inclination WSE flux rope or a low-inclination NWS flux rope. The \textit{in situ} data, again, show a clear flux rope identified from enhanced magnetic field magnitude and smooth field rotation. The MVA yields an axis tilt of $10^{\circ}$ and left-handed chirality. Hence, the \textit{in situ} flux rope clearly has a low-inclination and is of type NWS. If we also consider as a match cases where the flux rope is of intermediate type (\textit{i.e.} close to $45^{\circ}$ inclination at the Sun and/or \textit{in situ}), then the flux rope types agree between the Sun and \textit{in situ} for 11 (55\%) analysed events.

\begin{figure}[ht]
\includegraphics[width=.99\textwidth]{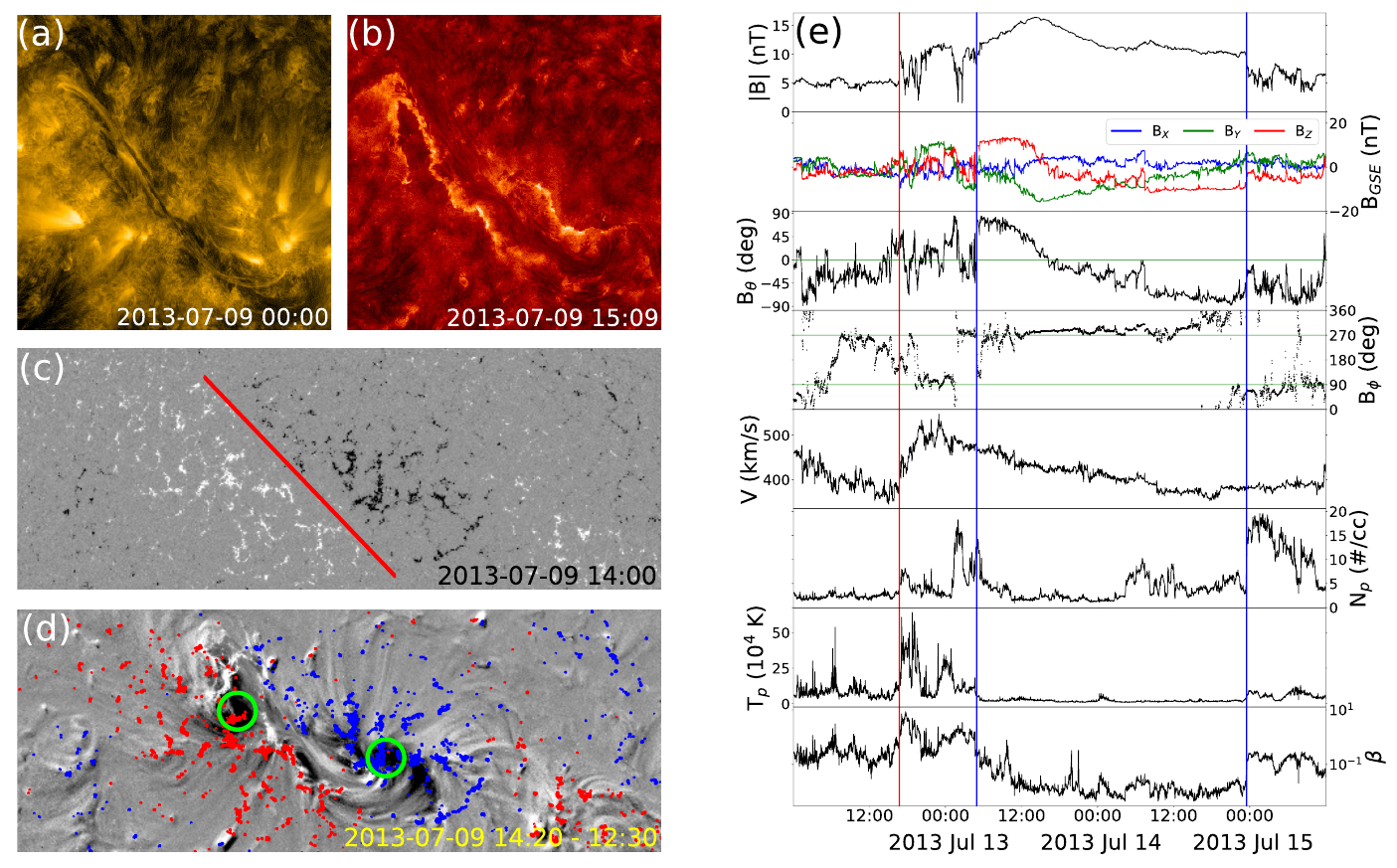}
\caption{Event 18, which is intermediate between a WSE- and a NWS-type at the Sun and is a NWS-type \textit{in situ}. (a) The reverse-S filament shape seen by SDO/AIA 171 \AA \, that indicates left-handed chirality. (b) Reverse-J shaped flare ribbons as seen in 304 \AA, a sign of a left-handed flux rope. (c) SDO/HMI magnetogram (saturated at $\pm 200$ G) showing the PIL approximated as a straight line (in red). (d) Base-difference AIA image in 211 \AA \, saturated at $\pm 200$ DN s$^{-1}$ pixel$^{-1}$ and overlaid with HMI magnetogram contours saturated at $\pm 200$ G (blue = negative polarity, red = positive polarity). The dimming regions (signatures of the flux rope footpoints) have been circled in green. (e) The ICME as observed in situ by \textit{Wind} (see Figure \ref{fig:event10} for details).}
\label{fig:event18}
\end{figure}

A clear example of a case where the flux rope types at the Sun and \textit{in situ} do not match is Event 17 (Figure \ref{fig:event17}). According to our analysis of the near-Sun observations, the intrinsic flux rope type is in the intermediate state between a high-inclination WSE-type and a low-inclination NWS-type. The helicity proxies that we used for this event were a clear reverse-S sigmoid (Figure \ref{fig:event17}a), a left-handed crossing of filament threads (Figure \ref{fig:event17}b), and reverse-J flare ribbons (visible in Figure \ref{fig:event17}d). The tilt angle at the Sun was estimated to be $55^{\circ}$. In this case, the tilt angle was deduced both from the PEAs seen in EUV (Figure \ref{fig:event17}c) and the orientation of the PIL (not shown). Visual inspection of the \textit{in situ} measurements, however, shows a strongly northward field during the passage of the entire ICME, and suggests that the flux rope type is ENW. MVA yields a high-inclination flux rope with a tilt of $-62^{\circ}$, in agreement with the visual analysis. This means that the axis orientation changed by $\sim 180^{\circ}$ from the Sun to L1.

\begin{figure}[ht]
\includegraphics[width=.99\textwidth]{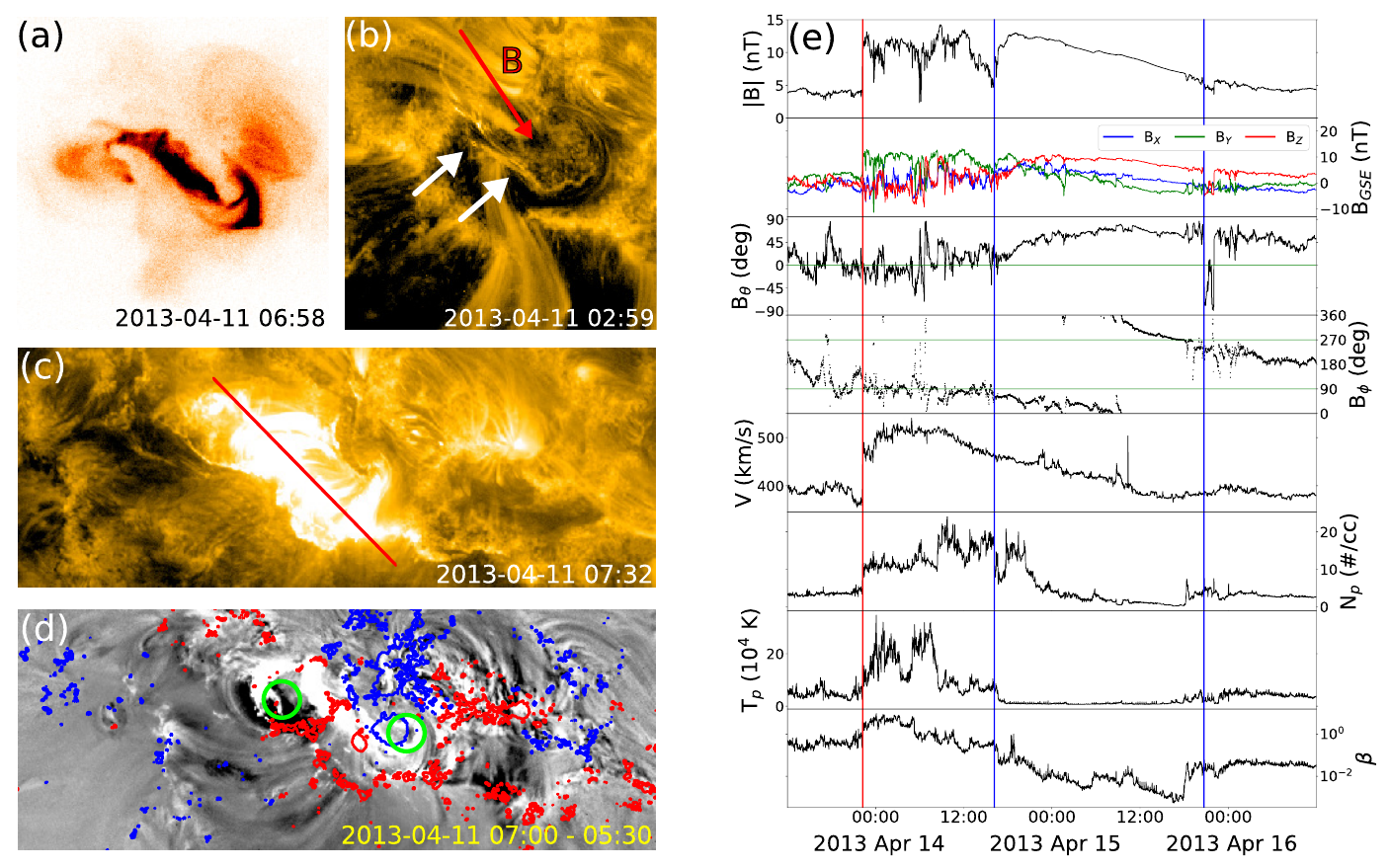}
\caption{Event 17, which is intermediate between a WSE- and a NWS-type at the Sun and is an ENW-type \textit{in situ}. (a) Reverse-colour soft X-ray images taken with \textit{Hinode}/XRT, showing an erupting reverse-S sigmoid, indicative of a left-handed flux rope. Filter wheel 1 is in the ``Beryllium thin'' (Be thin) position, while filter wheel 2 is Open. (b) Left-handed crossings of filament threads (indicated by the white arrows) as seen by SDO/AIA in 171  \AA . The direction of the magnetic field along the filament is also shown (in red). (c) 171  \AA \, observations showing the PEAs approximated as a straight line (in red). (d) Base-difference AIA image in 211 \AA \, saturated at $\pm 400$ DN s$^{-1}$ pixel$^{-1}$ and overlaid with HMI magnetogram contours saturated at $\pm 200$ G (blue = negative polarity, red = positive polarity). The dimming regions inside the reverse-J shapes of the flare ribbons (signatures of the flux rope footpoints) have been circled in green. (e) The ICME as observed in situ by \textit{Wind} (see Figure \ref{fig:event10} for details).}
\label{fig:event17}
\end{figure}

We also note that for two events (Events 12 and 20) the axis orientation resulting from MVA did not agree with our visual determination. Event 12 is clearly a case where the flux rope crosses \textit{Wind} far from its centre; MVA does not perform well for such events. However for Event 20, it is not obvious why MVA yields a low-inclination flux rope ($\theta_{\text{MVA}}=30^{\circ}$), while observations suggests an intermediate event. Anyhow, the flux ropes types would not match between the Sun and L1, as the possible flux rope types \textit{in situ} would be SWN and ESW. 

The minimum \textit{Dst} value for each analysed CME is reported in Table \ref{tab:ICME_types}. We note that five (25\%) CMEs caused minor or no storm (\textit{i.e.}, \textit{Dst}$_{\text{min}}$ > -50 nT), 11 (55\%) caused a moderate storm (-50 nT >  \textit{Dst}$_{\text{min}}$ > -100 nT), and four (20\%) caused an intense storm (\textit{Dst}$_{\text{min}}$ < -100 nT). The six high-inclination flux ropes detected \textit{in situ} with a southward axial field (\textit{i.e.}, of types ESW and WSE) all produced at least a moderate storm, and three of them produced intense storms. This is expected, since the primary requirement for a geomagnetic storm is that the interplanetary magnetic field is southward for a sufficiently long period of time. In total, our data set included five high-inclination and two ``intermediate'' ICMEs with northward axial fields. Four of these corresponded to minor or no storm (\textit{i.e.}, \textit{Dst}$_{\text{min}}$ > -50 nT), but two (Events 4 and 8) caused moderate storms and one (Event 5), an intense storm. In these three events, \textit{Dst}$_{\text{min}}$ was reached either before or shortly after (within four hours of) the passage of the ICME leading edge over L1. This suggests that these storms were driven by the sheath ahead of the ICME. A significant fraction of magnetic storms are, in fact, purely sheath-driven \citep{tsurutani1988,huttunen2002,huttunen2004, siscoe2007, kilpua2017a}. The sheaths of these three events, indeed, featured periods of strong southward fields (\textit{i.e.}, $B_{Z} \leq -10$ nT).

Figure \ref{fig:angles} provides a visual representation of the results reported in Tables \ref{tab:CME_types} and \ref{tab:ICME_types}, by comparing the flux rope clock angles at the Sun to those at L1. The figure highlights how the expected flux rope type at Earth can change due to rotation of the flux rope axis in the corona or in interplanetary space. The events are grouped according to their chirality, in order to look for possible patterns that might be related to the sign of the helicity \citep[\textit{i.e.}, clockwise rotation is expected for right-handed chirality and counterclockwise rotation for left-handed chirality,][]{fan2003,green2007,lynch2009}. We note from Figure \ref{fig:angles} an obvious pattern: the axis clock angles at the Sun are clustered in the vicinity of the dashed lines both for left- and right-handed flux ropes (\textit{i.e.}, they lie along the Northwest--Southeast diagonal for right-handed events and the Northeast--Southwest diagonal for left-handed events). A similar pattern was found by \citet{marubashi2015}. The clock angle change from the Sun to Earth is < $90^{\circ}$ for 13 (65\%) events. 

The remaining seven (35\%) events experienced > $90^{\circ}$ rotation of their central axis. Of these, one event (Event 2) experienced an apparent rotation of its axis by $\sim 100^{\circ}$, while the other six (30\%) seemed to rotate by $\gtrsim 120^{\circ}$. Of these latter six cases, three events are right-handed and three events are left-handed. All of them were formed in active regions. Such large rotations have been reported previously in the literature \citep[e.g.,][]{harra2007,kilpua2009}. We have not considered here how the flux rope chirality affects the sense of rotation of the clock angle, because, in some cases, the MVA can have large errors related to the \textit{in situ} clock angle (up to about $\pm 90^{\circ}$ when the flux rope is crossed very far from its central axis) and because, from a forecasting perspective, it is more useful to consider the smallest rotation angle between the two orientations (\textit{i.e.}, < $\pm 180^{\circ}$).  

We remark that a large fraction of events had their solar tilt angle close to $45^{\circ}$. In this regard, we point out that, when the flux rope axis orientation determined from solar observations is close to the intermediate one, the expected flux rope type at Earth can change even due to a relatively small amount of rotation ($\sim 20^{\circ}$). 

\begin{figure}[ht]
\centering
\includegraphics[width=.99\linewidth]{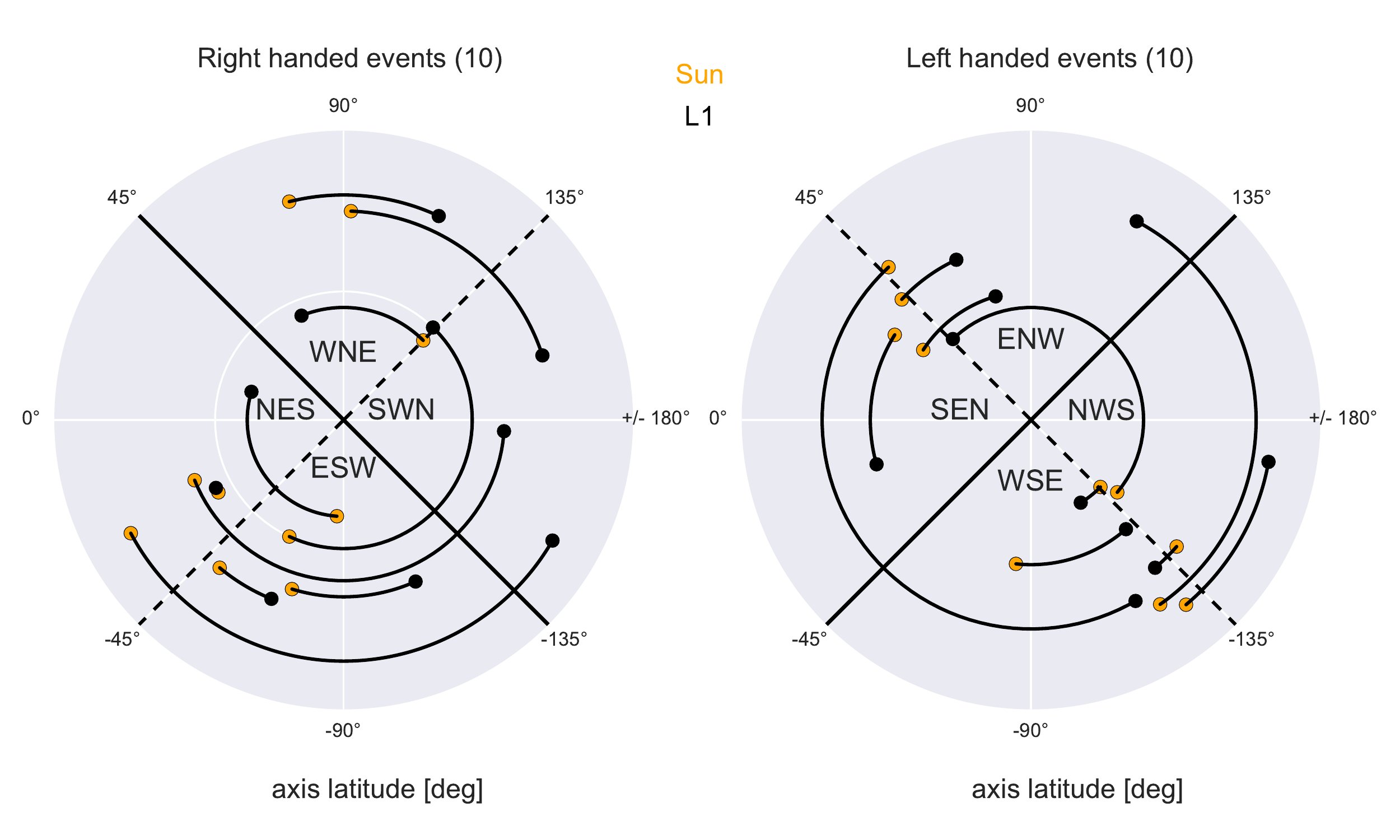}
\caption{Change in the flux rope clock angle from the Sun to L1, split into right- and left-handed events. The yellow dots represent the flux rope axis orientation at the Sun (the average between the orientations of the PIL and the PEAs), while the black dots indicate the orientation at L1 (taken from the axis orientation resulting from the MVA). Rotations are assumed to be < $180^{\circ}$ , \textit{i.e.} clockwise and counterclockwise rotations depending on chirality are not considered. Error bars are not included in the plot, but we assume that the error for the solar orientations can be up to $\pm 20^{\circ}$ and for the \textit{in situ} one can be up to $\pm 45^{\circ}$.}
\label{fig:angles}
\end{figure}

It is also interesting to investigate whether the CME source region location or the crossing distance of the spacecraft along and across the ICME affect whether the intrinsic and \textit{in situ} flux rope types match. Figure \ref{fig:sources} shows the source coordinates of the CMEs, measured as the mid point between the flux rope footpoints. The colors show whether the intrinsic and \textit{in situ} flux ropes matched or not and the symbols give an estimate of the crossing distance from the ICME axis (Figure \ref{fig:sources}a) and the ICME nose (Figure \ref{fig:sources}b). We remind that the crossing distance across the flux rope was estimated through the ratio $\langle|B_{\min}|\rangle/\langle B\rangle$ in the MVA reference system, while the crossing distance along the flux rope was estimated through the location angle (see Section \ref{subsec:insitu}). It is clear that there is no obvious pattern, regarding either the source location or the crossing distance from the axis and nose of the ICMEs. Nearly all source regions are clustered relatively close to the solar disc centre, within $\pm 30^{\circ}$ both in latitude and longitude. The events with the largest distances from the disc centre are, however, identified as mismatches or intermediate cases.  

\begin{figure}[ht]
\includegraphics[width=.99\textwidth]{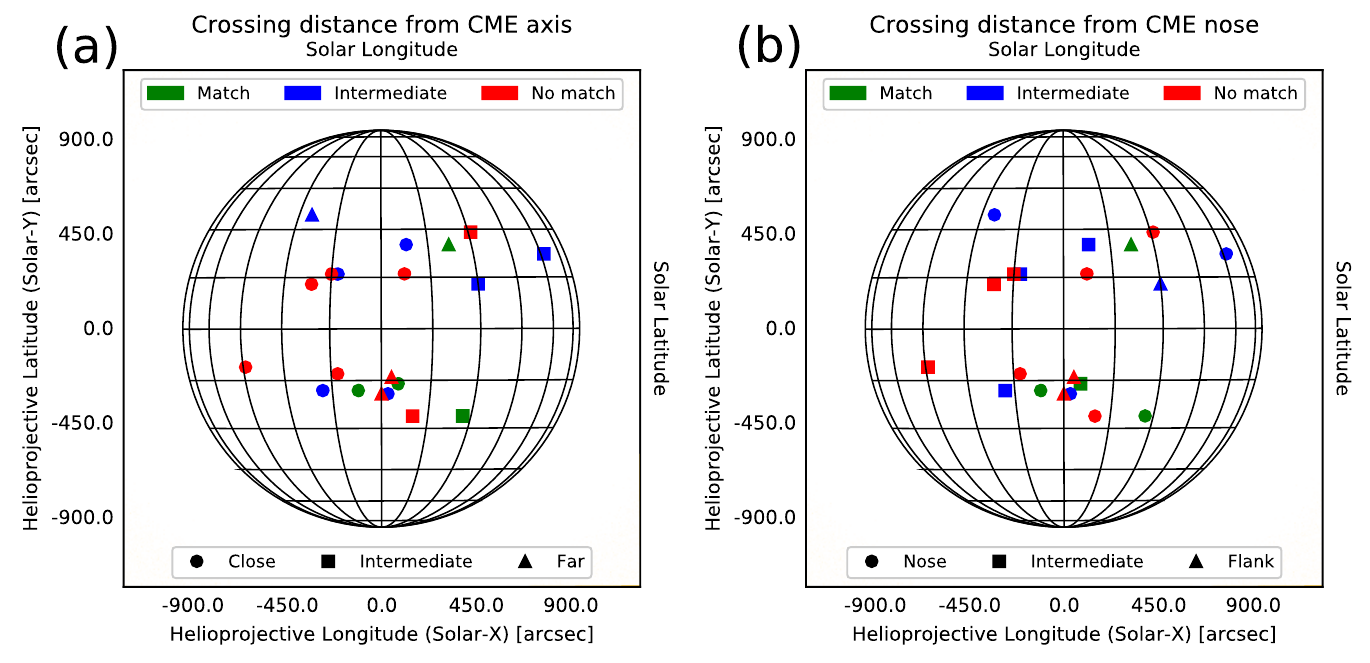}
\caption{Location of the source regions of the 20 CMEs under analysis. The different colours refer to how the flux rope types match between the Sun and L1: exact match (green), intermediate match (blue), and no match (red). The different symbols refer to the spacecraft crossing distance along and across the ICME. For panel (a), crossing closer to the axis (circles, $\langle|B_{\min}|\rangle/\langle B\rangle$ < 0.2), intermediate crossing (squares, 0.2 < $\langle|B_{\min}|\rangle/\langle B\rangle$ < 0.4), and crossing farther from the axis (triangles, $\langle|B_{\min}|\rangle/\langle B\rangle$ > 0.4). For panel (b), nose crossing (circles, $|\text{L}|$ < $15^{\circ}$), intermediate crossing (squares, $15^{\circ}$ < $|\text{L}|$ < $30^{\circ}$), and crossing closer to the flank (triangles, $|\text{L}|$ > $30^{\circ}$).}
\label{fig:sources}
\end{figure}

\section{Discussion and Conclusions}

In this work, we have analysed 20 CME events that had a clear and unique connection from the Sun to Earth as determined by heliospheric imaging. We have analysed their magnetic structure (specifically flux rope type) both at the Sun and \textit{in situ} at the Lagrange L1 point. The analysis of the solar sources was performed following the scheme presented in \citet{palmerio2017}. In particular, several multiwavelength indirect proxies were used to obtain the flux rope helicity sign (chirality), the axis tilt, and the direction of the magnetic field at the central axis, in order to determine the flux rope type of the erupting CME. The \textit{in situ} flux rope type was determined by visual inspection of magnetic field data and by applying the MVA technique.

One important work towards understanding of the magnetic structure of ICMEs with a flux rope structure and their solar counterparts was performed by \citet{bothmer1998}. The authors estimated the flux rope type of 46 ICMEs and found a unique association for nine ICMEs with quiet-Sun filament eruptions. In eight of the nine cases, they found agreement between the solar and \textit{in situ} flux rope types, where the intrinsic flux rope configuration was inferred from the orientation of the filament axis and its magnetic polarity, and the heliospheric helicity rule. A more recent study by \citet{savani2015} studied eight CME events from the Sun to Earth, using the \citet{bothmer1998} scheme to estimate the intrinsic flux rope configuration, and proved that the initial flux rope structure must be adjusted for cases originating from between two active regions. Indeed, our present study shows that the \citet{palmerio2017} scheme to determine the intrinsic flux rope type is applicable to several different types of CME eruptions. Our analysis included CMEs originating from a single active region, from pairs of nearby active regions, and from filaments located on the quiet Sun. The scheme succeeded in estimating the intrinsic flux rope type also for CME source regions that did not follow the hemispheric helicity rule. We remark that the chirality has been determined from observations rather than from applying the statistical helicity rule. The proxies that we used and their success rate (\textit{i.e.}, the percentage of the events to which we could apply them) are: arcade skew (90\%), S-shaped features (70\%), filament characteristics (60\%), flare ribbons (55\%), and magnetic tongues (15\%). We point out that, for the quiet Sun filaments, we were typically able to study filament characteristics only using H$\alpha$, while for active regions filaments, we typically used EUV observations. The flux rope axis orientation at the Sun could by determined both from PIL and PEAs in 65\% of cases and from PIL only in rest of the cases. 

We found that the flux rope types at the Sun (\textit{i.e.}, the intrinsic flux rope type) and \textit{in situ} matched only for four (20\%) events but, if intermediate cases are considered as a match, then the rate is considerably higher, 11 events (55\%). The tendency of the tilt of the flux rope axis at the Sun to be close to $45^{\circ}$ is hence problematic for determining between the eight traditional flux rope categories. As mentioned in Section \ref{sec:results}, this trend was noted by \citet{marubashi2015}. There is a tendency for bipolar active regions to emerge with a systematic deviation from the East--West direction, with the leading sunspot being closer to the solar equator. This pattern is known as Joy's law \citep{hale1919}. The tilt angle of bipolar sunspot groups (\textit{i.e.}, the line that connects two sunspots), however, tends to have an inclination of $1^{\circ}$--$10^{\circ}$ only due to Joy's law \citep[\textit{e.g.},][]{vandriel2015}. This means that the angle of the corresponding PIL tends to be $89^{\circ}$--$80^{\circ}$ tilted to the ecliptic upon emergence. Most of the PILs under analysis were clustered around $45^{\circ}$ tilt, which means that Joy's law cannot explain such tendency. Since magnetic tongues could be used as a helicity proxy for three events only (out of 14 CMEs originating from a single active region), then it follows that most of the studied active regions were in their decay phase. A possible cause for the PILs to increasingly change their alignment from North--South to Northwest--Southeast (Northeast--Southwest) for right-handed (left-handed) active regions is the Sun's differential rotation, which progressively acts on the PILs' tilt angle. This would also hold for active regions that are at the final phase of their decay, which are usually source regions for quite-Sun filament eruptions.

The frequent mismatch in flux rope type between the Sun and Earth suggests significant evolution after the eruption, particularly in terms of flux rope rotation. The comparison of the flux rope axis direction at the Sun and the Earth showed that for 35\% of the events that we studied (7 events) the difference between the axis directions at the Sun and \textit{in situ} was > $90^{\circ}$, with 20\% (\textit{i.e.}, 4 events) undergoing over $150^{\circ}$ rotation of their axis. All of the events that experienced a very large difference in the flux rope axis orientation originated from an active region. For the rest of the events (65\%; 13 events) the rotation was < $90^{\circ}$, and for 25\% of the events (\textit{i.e.}, 5 events) the difference was < $30^{\circ}$. Moreover, the four events that originated from a quiet-Sun filament seemed to rotate < $45^{\circ}$. This is in agreement with \citet{bothmer1998}, that found consistency in the flux rope configuration of erupting quiet-Sun filaments with their \textit{in situ} counterparts for eight out of nine cases. We therefore suggest that our lower percentage of matches between solar and \textit{in situ} flux rope types derives from the fact that we considered mostly active region CMEs in our dataset. We also showed that, at least for our relatively small data set, the difference between the axis orientations at the Sun and L1 did not seem to be obviously affected by the CME source location or by the crossing distance along and across the flux rope loop (Figure \ref{fig:sources}). We remind the reader that, in this analysis, we did not consider the expected sense of rotation dictated by the flux rope chirality, \textit{i.e.} clockwise (anticlockwise) for right- (left-) handed events. In fact, if we consider the smallest angle between the solar and \textit{in situ} flux rope orientations, then only ten events (50\%) seem to follow the sense of rotation expected from their chirality. This may either be because the remaining ten CMEs actually rotated in the opposite sense or that there was an external factor that counteracted the expected sense of rotation.

However, it is important to remark that the resulting flux rope orientation \textit{in situ} may depend on the fitting technique. \citet{alhaddad2013} analysed 59 ICMEs using four different reconstruction or fitting methods, and found that for one event only all four methods found an orientation of the ICME axis within $\pm 45^{\circ}$. Reconstructions done with different techniques usually disagree, and that has to be taken into account when comparing solar and \textit{in situ} orientations, especially when considering the sense of rotation of the axis for the low rotation cases. If we consider, \textit{e.g.}, only the cases that present a > $45^{\circ}$ angular difference (\textit{i.e.}, 11 events in total), then four (five) right-handed (left-handed) flux ropes seemed to rotate anticlockwise and two (zero) clockwise. The left-handed events, hence, seem all to follow the expected sense of rotation if the analysis is restricted to the large rotation cases.

It is noteworthy that the direct comparison between intrinsic and \textit{in situ} flux rope types can be performed only for a fraction of all CME--ICME pairs. As discussed in Section 2, we considered 47 candidates from the LINKCAT catalogue and ended up with only 12 events. The problems are related to (1) correctly connecting the CME--ICME pair, (2) excluding interacting events, and (3) the requirement for the relevant observations to be sufficiently clear both at the Sun and \textit{in situ} in order to estimate the flux rope type. In particular, many ICMEs do not show clear enough rotation of the field to determine the flux rope type. At the Sun, some CMEs may be so-called stealth CMEs \citep[\textit{e.g.},][]{robbrecht2009,kilpua2014,nitta2017}, \textit{i.e.} they lack obvious disk signatures, or have curved PEAs and/or PIL so reliable determination of the axis orientation is not possible. However, the cases for which determination of the intrinsic and \textit{in situ} flux rope types is possible are often geoeffective, as they show clear magnetic field enhancements and organized rotation of the magnetic field. In addition, as remarked in the Introduction, one important point to keep in mind for real-time space weather forecasts is that it is often difficult to predict if an erupting CME would impact Earth at all. Hence, a further investigation to study the applicability of the methods described in this article for forecasting would require to start at the Sun without first identifying CME--ICME pairs.

As already mentioned in the Introduction, determination of the intrinsic flux rope type is a crucial step in space weather forecasting (as the input to different models), and as showed in this paper, in a fraction of cases it gives a good estimate of the flux rope magnetic structure at L1. Our results, however, strongly highlight the importance of capturing the amount of rotation and/or distortion that the flux rope experiences in the corona and in interplanetary space. This was stated already in the work by \citet{savani2015}, that highlights the importance of including evolutionary estimates of CMEs from remote sensing for space weather forecasts. The flux rope axis direction \textit{in situ} can be, \textit{e.g.}, estimated by considering coronagraph data in addition to solar disc observations \citep{savani2015}. Concerning flux rope rotations, in fact, several studies suggest that the most dramatic rotation occurs during the first few solar radii of a CME's propagation \citep[\textit{e.g.},][]{vourlidas2011,isavnin2014,kay2016}. Indeed, rotation can also occur even during the eruption \citep[\textit{e.g.},][]{green2007,lynch2009,bemporad2011,thompson2012}.

Finally, we remark that \textit{in situ} data are one-dimensional and that a single spacecraft's trajectory through a CME may not reflect the global shape and orientation of the flux rope. The flux rope type that is seen at Earth may depend on where the spacecraft crosses the ICME (\textit{i.e.} the crossing distance from the ICME axis, named the impact parameter, and/or from the ICME nose) and on local distortions that might be present within an ICME. In terms of the latter, \citet{bothmer2017} recently demonstrated that kinks present in the CME source region seem to be reflected in the erupting flux rope during its expansion and propagation. \citet{owens2017} also showed that CMEs cease to be coherent magnetohydrodynamic structures within 0.3 AU of the Sun, and that their appearance beyond this distance is that of a dust cloud. This means that local deformations that may arise during the CME propagation do not propagate throughout the whole CME body. Nevertheless, the space weather effects at Earth depend strongly on the magnetic structure that is measured at L1, meaning that a significant step towards the improvement of current space weather forecasting capabilities is the prediction of the flux rope axis rotation (whether proper or apparent) during propagation. Other important factors to take into account for future space weather predictions are the crossing location, both along and across the flux rope, and eventual local distortions of the CME body.

\acknowledgments
EP acknowledges the Doctoral Programme in Particle Physics and Universe Sciences (PAPU) at the University of Helsinki. This project has received funding from the European Research Council (ERC) under the European Union's Horizon 2020 research and innovation programme (grant agreement n$^{\circ}$ 4100103). EK also acknowledges UH Three Year Grant project 75283109. CM's work was supported by the Austrian Science Fund (FWF): [P26174-N27]. AJ and LG acknowledge the support of the Leverhulme Trust Research Project Grant 2014-051. LG also acknowledges support through a Royal Society University Research Fellowship. We also thank the two anonymous reviewers, whose suggestions have significantly improved this article.

We thank the HELCATS project, that has received funding from the European Union's Seventh Framework Programme for research, technological development and demonstration under grant agreement no 606692. This research has made use of SunPy, an open-source and free community-developed solar data analysis package written in Python \citep{sunpy2015}, and the ESA JHelioviewer software. We thank the geomagnetic observatories (Kakioka [JMA], Honolulu and San Juan [USGS], Hermanus [RSA], INTERMAGNET, and many others) for their cooperation to make the provisional and the final \textit{Dst} indices available.

\section*{Sources of Data and Supplementary Material}
\label{sec:data}
\noindent Catalogues:\\~\\
LINKCAT, doi:10.6084/m9.figshare.4588330.v2,\\ \url{https://doi.org/10.6084/m9.figshare.4588330.v2}\\
ARRCAT, doi:10.6084/m9.figshare.4588324.v1,\\
\url{https://doi.org/10.6084/m9.figshare.4588324.v1}\\~\\
\noindent ICME Lists:\\~\\
Near-Earth Interplanetary Coronal Mass Ejections List, Richardson, I., and Cane, H.,\\
\url{http://www.srl.caltech.edu/ACE/ASC/DATA/level3/icmetable2.htm}\\
Wind ICME List, Nieves-Chinchilla, T., \textit{et al.},\\ \url{https://wind.nasa.gov/ICMEindex.php}

\listofchanges

\end{document}